\documentclass[aps,twocolumn,preprintnumbers,prd,superscriptaddress,nofootinbib,10pt]{revtex4-2}

\usepackage{aas_macros}
\usepackage{graphicx,epsfig}
\usepackage{amsmath,amssymb}
\usepackage{lineno}
\usepackage{subfigure}
\usepackage{xcolor}
\usepackage[english]{babel}

\usepackage{comment}
\usepackage{multirow}
\usepackage[normalem]{ulem}
\usepackage{float}

\usepackage{hyperref}

\hypersetup{
 colorlinks=true,
 linktoc=all,
 linkcolor=red,
 citecolor=blue,
 urlcolor = blue}

\newcommand{\ii}{\mathrm{i}} 
\newcommand{\dd}{\mathrm{d}} 

\newcommand{\bfxi}{\mbox{\boldmath{$\xi$}}}
\newcommand{\bfeta}{\mbox{\boldmath{$\eta$}}}

\newcommand{\beq}{\begin{equation}}
\newcommand{\beqa}{\begin{eqnarray}}
\newcommand{\eeq}{\end{equation}}
\newcommand{\eeqa}{\end{eqnarray}}

\begin{document}

\title{Parameter estimation of microlensed gravitational waves with Conditional Variational Autoencoders}

\author{Roberto Bada Nerin}
\email{roberto.bada-nerin@ligo.org}
\affiliation{Institut de Ci\`encies del Cosmos (ICCUB), Facultat de F\'{i}sica, Universitat de Barcelona, Mart\'i i Franqu\`es 1, E-08028 Barcelona, Spain}

\author{Oleg Bulashenko}
\email{oleg@fqa.ub.edu}
\affiliation{Institut de Ci\`encies del Cosmos (ICCUB), Facultat de F\'{i}sica, Universitat de Barcelona, Mart\'i i Franqu\`es 1, E-08028 Barcelona, Spain}

\author{Osvaldo~Gramaxo Freitas}
\affiliation{Centro de F\'{\i}sica das Universidades do Minho e do Porto (CF-UM-UP), Universidade do Minho, 4710-057 Braga, Portugal}
\affiliation{Departamento de Astronom\'{\i}a y Astrof\'{\i}sica, Universitat de Val\`encia, Dr. Moliner 50, 46100, Burjassot (Val\`encia), Spain}
\email{ogf1996@gmail.com}

\author{Jos\'e~A.~Font}
\affiliation{Departamento de Astronom\'{\i}a y Astrof\'{\i}sica, Universitat de Val\`encia, Dr. Moliner 50, 46100, Burjassot (Val\`encia), Spain}
\affiliation{Observatori Astron\`omic, Universitat de Val\`encia,  Catedr\'atico Jos\'e Beltr\'an 2, 46980, Paterna (Val\`encia), Spain}
\email{j.antonio.font@uv.es}

\begin{abstract}
    Gravitational lensing of gravitational waves (GWs) provides a unique opportunity to study cosmology and astrophysics at multiple scales. Detecting microlensing signatures, in particular, requires efficient parameter estimation methods due to the high computational cost of traditional Bayesian inference. In this paper we explore the use of deep learning, namely Conditional Variational Autoencoders (CVAE), to estimate parameters of microlensed binary black hole (simulated) waveforms. We find that our CVAE model yields accurate parameter estimation and  significant computational savings  compared to Bayesian methods such as \textsc{bilby} (up to five orders of magnitude faster inferences).
Moreover, the incorporation of CVAE-generated priors into \textsc{bilby}, based on the 95\% confidence intervals of the CVAE posterior for the lensing parameters, reduces Bilby's average runtime by around 48\% without any penalty on accuracy. Our results suggest that a CVAE model is a promising tool for future low-latency searches of lensed signals. Further applications to actual signals and integration with advanced pipelines could help extend the capabilities of GW observatories in detecting microlensing events.
\end{abstract}

\date{\today}

\maketitle


\section{Introduction}


  The first three observing runs of the network of GW detectors Advanced LIGO~\cite{Aasi:2014jea}, Advanced Virgo~\cite{Acernese:2014hva} and KAGRA~\cite{KAGRA:2020tym} have led to the detection of 90 compact binary coalescence (CBC) events~\cite{GWTC1, GWTC2, GWTC2.1, GWTC3}. Their analysis is impacting different fields of research, e.g.~astronomy, astrophysics, cosmology, and fundamental physics, in a significant way. In particular the wealth of new data has provided updated information on the CBC merger rate and has allowed for new studies of populations of compact objects (specifically regarding the distribution of black hole masses) and for new tests of general relativity~\cite{GWTC-3_Population_Properties,LIGOScientific:2020tif}. One distinct aspect being scrutinized is the possibility of finding evidence of gravitational lensing on the observed GW signals~\cite{lensingO3a,lensingO3b}. If it took place, lensing by massive objects along the line of sight between the detector and the source might significantly distort the signal leading to e.g.~beating patterns, signal amplitude changes, signal repetition,  or a combination of these phenomena. 
  
  The LIGO-Virgo-KAGRA (LVK) Collaboration has conducted several searches for GW lensing, including both strong lensing and microlensing signatures. 
  The most recent searches have accounted for all signals from the whole sample of binary black hole (BBH) mergers reported in the third observing run~\cite{lensingO3a,lensingO3b}. None of the searches conducted in those studies has yielded conclusive evidence for GW lensing~\cite{lensingO3b, janquart-23}. Other groups, however, have put forward different proposals suggesting that the high masses observed in some BBH systems could be explained by the magnification effect of gravitational lensing \cite{diego-21,bianconi-22,canevarolo-24} (see also related discussions in~\cite{lensingO3a,janquart-GW230529}). 

Although GW lensing has yet to be conclusively observed, it is widely recognized as a promising tool for GW astrophysics, offering significant potential for groundbreaking discoveries. This potential has been underscored in studies involving both third-generation terrestrial GW observatories \cite{Grespan2023,Lemon2024} and the space-based LISA mission~\cite{auclair-23}.
Certain gravitational lenses might be composed of objects that neither emit light nor neutrinos, making them undetectable through conventional means. Observing GW lensing could hence provide crucial insights into populations of elusive or poorly understood objects, such as intermediate-mass black holes \cite{lai18,gais22,meena24}, dark matter haloes \cite{choi21,gao21,guo22}, cosmic strings \cite{suyama06,pla-string16,pla-fresnel17}, or even potentially reveal entirely new classes of objects. The first detection of gravitational lensing by such phenomena would provide direct evidence of compact dark matter objects, offering a groundbreaking advance in astrophysics~\cite{jung19,liu19,choi21,wang21,urrutia21,gao21,basak21,fairbairn22,tambalo22,caliscan23,cheung24,jana24}.

Lensing effects on GWs, much like their electromagnetic counterparts, can manifest in a variety of regimes, each with distinct characteristics and implications. In the {\em strong lensing} regime, massive structures such as galaxies curve spacetime enough to create multiple images of the source. These repeated GW events are typically separated by time delays ranging from minutes to months \cite{sereno10,li18,oguri19}.
In contrast, the {\em microlensing} regime arises from smaller-scale objects like stars or stellar-mass compact objects, where the time delays are much shorter, typically spanning milliseconds to seconds
\cite{nakamura98,nakamura99,takahashi03,matsunaga06,cao14,christian18,diego19,liao19,orazio19,hou19,cheung20,cremonese21a,dalang21,chung21,yu21,seo21,cremonese21b,biesiada21,suvorov21,wright21,yeung21,BU-JCAP-21, caliscan22,ali22,tambalo22a,bondarescu22,LiuA-Kim23,LiuA-Wong23, urrutia23,savastano23,mishra23a,mishra23b,deka24,wright24,chen24,villarrubia24,chakraborty2410,LiuA-2410}.
This regime can amplify signals or imprint a distinctive 'beating' pattern in the waveform, caused by time delays that are comparable to the wave’s period\footnote{
Some authors further distinguish between microlensed GW signals, occurring in the wave optics regime, 
and millilensed GW signals, which fall within the geometric optics regime and are associated with comparatively higher lens masses \cite{LiuA-Kim23,LiuA-Wong23}.
Given that our research covers a broad range of masses encompassing both the wave-optics and geometric-optics limits (utilizing the hybrid transmission factor as detailed later), we will collectively refer to both regimes as 'microlensing' without distinction, consistent with the classification adopted in Refs.~\cite{lensingO3a,lensingO3b}.}.
Microlensing has profound astrophysical implications and has garnered considerable attention in GW astronomy, as it provides unique insights into small-scale structures in the universe. For example, lensing amplification can bias the inferred parameters of a source, such as its distance and mass, if left unaccounted for. A lensed GW may appear closer and more massive than it truly is, underscoring the importance of identifying lensing effects to avoid systematic errors.

In this paper, we contribute to these efforts by focusing on the microlensing regime and presenting results from a study based on the point-mass lens model. This model, the simplest and most illustrative for microlensing, captures lensing effects as a function of the redshifted lens mass $M_{Lz}$ and the source position parameter $y$, defined relative to the line of sight.

The straightforward approach to detecting microlensing is to do Bayesian parameter estimation of individual events with lensing added to the waveform model. Currently, the pipeline in charge of performing these searches is \textsc{Gravelamps}~\cite{wright21}. The biggest limitation of a Bayesian approach is computational efficiency, as it can be extremeley time consuming. 
Recent studies have demonstrated that deep-learning techniques offer a powerful way to perform fast parameter estimation in (unlensed) GW data 
(see the comprehensive review~\cite{cuoco24} and refrences therein).
In particular, Conditional Variational Autoencoders (CVAE) \cite{cvaes,Gabbard_2021}, 
normalizing flows \cite{Green_2020,Green_2021}, together with its extensions \textsc{Nessai} \cite{Williams_2021,Williams_2023},  \textsc{Dingo} \cite{Dax-21-DINGO,Dax-22-DINGO-IS} and flow matching \cite{Dax-23-flow-match}, have shown impressive potential in capturing complex features of GW signals, making them a valuable tool in contexts where speed is essential (e.g.~low-latency searches). 
Notably, CVAEs have already demonstrated their versatility in addressing a variety of GW analysis challenges, including continuous waves \cite{Bayley_2022}, ringdown signals \cite{Yamamoto-20}, and postmerger signals from binary neutron stars \cite{Whittaker-22}, among others.
Despite these advances, neither CVAEs nor normalizing flows have yet been applied to the parameter estimation of lensed GWs.
Other deep-learning approaches have been used for the  identification of microlensed \cite{kim20} 
and strong-lensed \cite{goyal21,magare24} signals, as well as on parameter estimation  and detection of strong lensing in the electromagnetic case~\cite{levasseur17,jacobs19}.  The aim of the present investigation is to bridge that gap, leveraging the unique capabilities of CVAEs to address the complex problem of parameter estimation in microlensing scenarios.

To do so we train a CVAE on a dataset of simulated, microlensed GW signals, enabling it to learn a probabilistic latent representation of the lensing parameters conditioned on the wave's time series. The results show that such a model could be deployed in inference tasks, offering a practical way to integrate microlensing models into real-time analysis frameworks, complementing traditional Bayesian parameter estimation methods, such as those based on packages like \textsc{Bilby} \cite{bilby_paper} or \textsc{Gravelamps}~\cite{wright21}. In particular, we demonstrate that incorporating CVAE-generated priors into \textsc{Bilby} significantly reduces its inference time, thereby enhancing the practicality of Bayesian methods for microlensing detection. 

This paper is organized as follows: In Section \ref{sec_pml} we review the theoretical framework in which microlensing and, in particular, the point-mass lens model, are based. Section \ref{sec_cvae} discusses the methodology employed in our study, this is, CVAEs. Next, in Section \ref{sec_data}, we present the different datasets used, together with some specifications on the data generation process. Section \ref{sec_results} presents the results of the training and testing processes, as well as a time comparison between CVAEs and \textsc{Bilby}. Finally, the conclusions of this work are presented in Section \ref{conclusions}. Specific details concerning the architecture of the model, the training and testing processes and the comparison with \textsc{Bilby} are reported in Appendix \ref{appendix}.

\section{Microlensing}
\label{sec_pml}
Any astrophysical object with sufficient mass can act as a lens for a passing GW.
Lensing effects occur when the source, the lens and the observer are all aligned within the Einstein angle $\theta_{\rm E}=R_{\rm E}/d_L$, i.e., the lens is located near the line of sight. 
The Einstein radius $R_{\rm E}$ is commonly expressed as:
\begin{equation}
    R_{\rm E} = \sqrt{\frac{4GM_L}{c^2}\, \frac{d_{LS}\,d_L}{d_S} },
\end{equation}
where $M_L$ is the mass of the lens, $d_{LS}$ is the angular diameter distance between the source and the lens, and $d_L$ and $d_S$ are the angular diameter distances to the lens and source at redshifts $z_L$ and $z_S$, respectively. 
$R_{\rm E}$ represents the characteristic length scale on the lens plane, is proportional to $\sqrt{M_L}$ and is typically much smaller than the cosmological distances $d_{LS}$, $d_L$, and $d_S$.  
This enables the lens mass to be projected onto a lens plane. In the thin-lens approximation, GWs propagate freely outside the lens, interacting only with a two-dimensional gravitational potential at the lens plane, where the lensing effect is ultimately captured in the transmission factor $F$ \cite{schneider-92}.

An unlensed GW signal from the source can be described by its frequency-domain strain $\tilde{h}(f)$, which is the Fourier transform of the time-domain strain $h(t)$.
After the signal passes through a lens, the resulting lensed waveform $\tilde{h}_L(f)$, which is ultimately detected, is the product of the transmission factor $F(f)$ and the original unlensed waveform:
\beq
    \tilde{h}_L(f) = F(f) \cdot \tilde{h}(f).
\label{lensed_strain}
\eeq
The transmission factor is determined by the Fresnel-Kirchhoff diffraction integral across the lens plane \cite{schneider-92}
\beq
F(f,\mathbf{y}) = -\ii f\, t_M
\iint e^{\ii \,2\pi f\, t_d (\mathbf{x},\mathbf{y})} \, \dd^2\mathbf{x},
\label{F_w}
\eeq
where the lensing time delay function is given by
\beq
t_d(\mathbf{x},\mathbf{y}) = t_M \, \left( \frac{1}{2}\, | \mathbf{x}-\mathbf{y} |^2
- \psi(\mathbf{x}) + \phi_m(\mathbf{y}) \right)
\label{t-delay}
\eeq
with the characteristic lensing time in physical units
\beq
t_M = (1+z_L) \,\frac{\xi_0^2\,d_S}{c\,d_L d_{LS}}.
\label{tM_xi}
\eeq
Here, $\xi_0$ is an arbitrary length scale in the lens plane, used to normalize both the impact parameter $\bfxi$ and the source position $\bfeta$ as follows \cite{takahashi03}:
$\mathbf{x}=\bfxi/\xi_0$, $\mathbf{y} = \bfeta \,d_L/(\xi_0 d_S)$.
In this notation, $\bfeta$ indicates the position of the source in the source plane, while $\mathbf{y}$ represents the projection of the source onto the lens plane, normalized by $\xi_0$.
For axially symmetric lenses, the vector $\mathbf{y}$ can always be reduced to the scalar $y$ by properly rotating the coordinate axes.
The factor $(1+z_L)$ is included to account for cosmological distance, where $z_L$ represents the redshift of the lens.
The properties of the lens are captured by the lensing potential $\psi(\mathbf{x})$. In the absence of lensing, where $\psi(\mathbf{x})=0$, the transmission factor \eqref{F_w}  simplifies to $|F|=1$,  leaving the waveform unaffected. 
For convenience, the phase $\phi_m(\mathbf{y})$ is defined such that the minimum of $t_d(\mathbf{x},\mathbf{y})$ for a fixed $\mathbf{y}$ is zero. 

Next, we will consider an isolated  point mass as the lens. In gravitational lensing, the point mass lens model (PML) is appropriate when the physical size of the lens is much smaller than the Einstein radius, as in the case of black holes, dense dark matter clumps, and similar compact objects. Due to its simplicity, the PML model has been widely used in the literature to interpret both electromagnetic \cite{deguchi86a,deguchi86b,schneider-92,petters-01} and GW lensing \cite{nakamura98,nakamura99,takahashi03,matsunaga06,cao14,christian18,diego19,liao19,orazio19,hou19,cheung20,cremonese21a,dalang21,chung21,yu21,seo21,cremonese21b,biesiada21,suvorov21,wright21,yeung21,BU-JCAP-21,caliscan22,ali22,tambalo22a,bondarescu22,urrutia23,savastano23,mishra23a,mishra23b,deka24,wright24,chen24,villarrubia24}. 

For the PML model, a natural choice for the length scale $\xi_0$ is the Einstein radius. With this normalization, the lensing potential is $\psi(\mathbf{x})=\ln{|\mathbf{x}|}$, for which the diffraction integral \eqref{F_w} can be solved analytically \cite{schneider-92,deguchi86a}
\begin{equation}
F = e^{\frac{1}{2}\pi^2 \nu} e^{\ii \pi \nu  \ln (\pi \nu)} \,
\Gamma ( 1- \ii \pi \nu ) \,_1F_1 ( \ii \pi \nu; \,1; \,\ii \pi \nu y^2 ) ,
\label{F_PML}
\end{equation}
where $\nu \equiv f t_M$ is the frequency of the GW $f$ scaled by the characteristic time $t_M$, $\Gamma(z)$ is the Gamma function, and $_1 F_1(a,b,z)$ is the confluent hypergeometric function\footnote{In our code we take the complex conjugate by replacing $\ii$ with $-\ii$ to align with the Fourier transform convention used in the Python libraries of the LVK Collaboration~\cite{lalsuite-fourier}.}.
With the chosen normalization $\xi_0=R_{\rm E}$, 
the time $t_M$, as expressed by Eq.~\eqref{tM_xi}, is proportional to the mass of the lens: 
\begin{equation}
t_M =2{\cal R_{\rm S}}/c \;\approx \; 1.97 \times 10^{-5} \,{\rm s}\; (M_{Lz}/M_\odot),
\label{tM}
\end{equation}
where ${\cal R_{\rm S}} =2GM_{Lz}/c^2$ denotes the Schwarzschild radius, and $M_{Lz}=M_L\,(1+z_L)$ is the  redshifted mass of the lens.

The absolute value of the transmission factor is obtained from Eq.~\eqref{F_PML} as follows \cite{deguchi86a}
\begin{eqnarray}
|F|=\displaystyle{
\left( \frac{2 \pi^2 \nu}{1-e^{-2 \pi^2 \nu}} \right)^{1/2}
 \left| _1F_1 ( \ii \pi \nu; \,1; \,\ii \pi \nu y^2 ) \right| }.
\label{eq:abs_F}
\end{eqnarray}
As seen from Eqs.~\eqref{F_PML} and \eqref{eq:abs_F}
the transmission factor, which is a function of frequency $f$, depends on two parameters:
(i) the mass of the lens $M_{Lz}$ through the time $t_M$
and (ii) the scaled offset of the source $y$.
Its behavior has already been described in the literature \cite{deguchi86a,takahashi03,matsunaga06} (in current notation, see also Refs.~\cite{BU-JCAP-21,bondarescu22}).

We are interested in the behavior of the transmission factor within the frequency range detectable by the LVK network, which spans approximately from 30 Hz to 1 kHz \cite{Abbott2020, Acernese_2015, PhysRevLett.123.231108, akutsu2020overviewkagradetectordesign, Buikema_2020, PhysRevLett.123.231107}.
For a given choice of parameters $M_{Lz}$ and $y$, the lensing effect on GWs in the LVK range can fall into the amplification region, the oscillating region, or an intermediate region, depending on the lens mass \cite{BU-JCAP-21}. For lower masses (e.g. $M_L = 30 M_\odot$), the primary effect is amplification. As the mass increases, oscillatory effects with a beating pattern emerge in the waveform \cite{bondarescu22}. 
At high frequencies, computing the transmission factor \eqref{eq:abs_F} in the oscillating region becomes numerically expensive. Fortunately, in this limit, the dominant contribution arises from two well-defined images of the source, corresponding to two stationary points of the time delay function. 
In general, the transmission factor in this geometrical optics limit can be written as a sum over all stationary points \cite{schneider-92,nakamura99,takahashi03},
\begin{equation}
F_{\rm GO}=\sum_j |\mu_j|^{1/2} e^{\ii(2\pi f \,t_d (\mathbf{x}_j,\mathbf{y})- n_j\pi/2)},
\label{F_GO}
\end{equation}
where 
$\mu_j=(\det \left(\partial \mathbf{y} / \partial \mathbf{x}_j 
\right))^{-1}$ is the magnification of the j-th image and $n_j=0,1,2$ corresponds to the Morse index for a minimum, saddle point and maximum of $t_d$, respectively.
The positions of the images $\mathbf{x}_j$ are determined by the lens equation, which for the PML model is given by $\mathbf{y}=\mathbf{x}-\mathbf{x}/|\mathbf{x}|^2$ \cite{schneider-92}.
For $\mathbf{y}=(y,0)$, with $y>0$, the two images on the lens plane $\mathbf{x}_{1,2}=(x_{1,2},0)$ are determined by
$x_{1,2}=(y\pm \sqrt{y^2+4})/2$, which leads to the following formula for the transmission factor \cite{takahashi03}:
\beq
F_{GO}= \left(\sqrt{\mu_+} + \sqrt{\mu_-} \,e^{2 \ii \alpha} \right) e^{\ii \phi_1},
\label{F_GO_PML}
\eeq
Here, the magnification for each image, expressed as $\mu_{\pm}=(v+v^{-1}\pm 2)/4$ with $v \equiv y/\sqrt{y^2+4}$, is ultimately a function of $y$ alone, while the phase 
$\alpha = \pi f t_M \tau_{21} - \pi/4$, where 
$\tau_{21} = 2v/(1-v^2) + \ln [(1+v)/(1-v)]$, 
also depends on frequency and lens mass. 
The function $t_M \tau_{21}$ represents the time delay between the two images, while $\phi_1$ denotes the phase of the first image, taken as a reference \cite{BU-JCAP-21}.

For practical purposes and to minimize excessive computational time, we employ a hybrid function for the transmission factor. 
This function utilizes the full wave $F$ from Eq.~\eqref{F_PML} at low frequencies and its GO limit \eqref{F_GO_PML} at high frequencies.
The simplicity of the GO formula significantly accelerates computations. To ensure a smooth transition between these two solutions, we choose the frequency of the third oscillation maximum, $f_{m}=2.25/(t_M\tau_{21})$, as the matching point (see Refs.~\cite{BU-JCAP-21,bondarescu22} for details).



\section{Conditional Variational autoencoder}
\label{sec_cvae}

Given a noisy GW waveform, $h$, with some parameters, $\Lambda$, the objective of Bayesian parameter estimation is to find the posterior probability distribution of the parameters conditioned by this $h$, $p(\Lambda|h)$. Following Bayes theorem, this is given by 
\begin{equation}
    p(\Lambda|h)\propto p(h|\Lambda)p(\Lambda),
\end{equation}
where $p(h|\Lambda)$ represents the likelihood and $p(\Lambda)$ the prior of the parameters. The common way of approaching this problem with Bayesian inference methods involves several explicit likelihood evaluations, which results in large computational times~\cite{Christensen_2022}. In this context, likelihood-free \cite{Cranmer20} machine learning methods have been proposed as a way of speeding up the process
\cite{cvaes,Gabbard_2021,Green_2020,Green_2021,Williams_2021,Williams_2023,Dax-21-DINGO,Dax-22-DINGO-IS,Dax-23-flow-match}.

A kind of models that have proved specially useful for the estimation of posterior probability distributions are CVAEs. Autoencoders \cite{aes} are a type of neural network used to learn efficient, compressed representations of data in a space -- the so-called latent space -- of lower dimension than the input. An autoencoder consists of two main parts:

\begin{enumerate}
    \item Encoder: Maps the input data into a lower-dimen\-sional space.
    \item Decoder: Reconstructs the original data from the compressed representation.
\end{enumerate}

The goal is to minimize the difference between the input and the output, thereby learning meaningful features of the data without supervision. Autoencoders are widely used in tasks like data compression, denoising, and anomaly detection  \cite{ae_review}. A special type of autoencoders are Variational Autoencoders (VAEs) \cite{vaes}. These learn probabilistic latent representations of data. This means that, unlike traditional autoencoders (which map the input to a fixed point in the latent space), VAEs model the latent space as a probability distribution, typically Gaussian. In other words, the features learnt are the distribution parameters -- the mean and the standard deviation in the case of a Gaussian. Then, the decoder reconstructs the data by sampling from this latent space. VAEs are often used in generative modeling, anomaly detection, and unsupervised learning \cite{ae_review, NELOY2024100572}. Finally, a CVAE \cite{cvaes} is an extension of the VAE that introduces conditioning information into the model. This conditioning could be labels, categories, or any side information related to the data. In CVAEs, the conditioning information is used both by the encoder and the decoder, for example, by concatenating it with the input data and with the sample taken from the latent space. In summary, the main difference from VAEs is that CVAEs generate outputs that are conditioned on specific information, making them useful for tasks like controlled data generation and structured prediction (e.g.~generating images of a specific class).

\begin{figure}[t]
    \centering
    \includegraphics[width=1\linewidth]{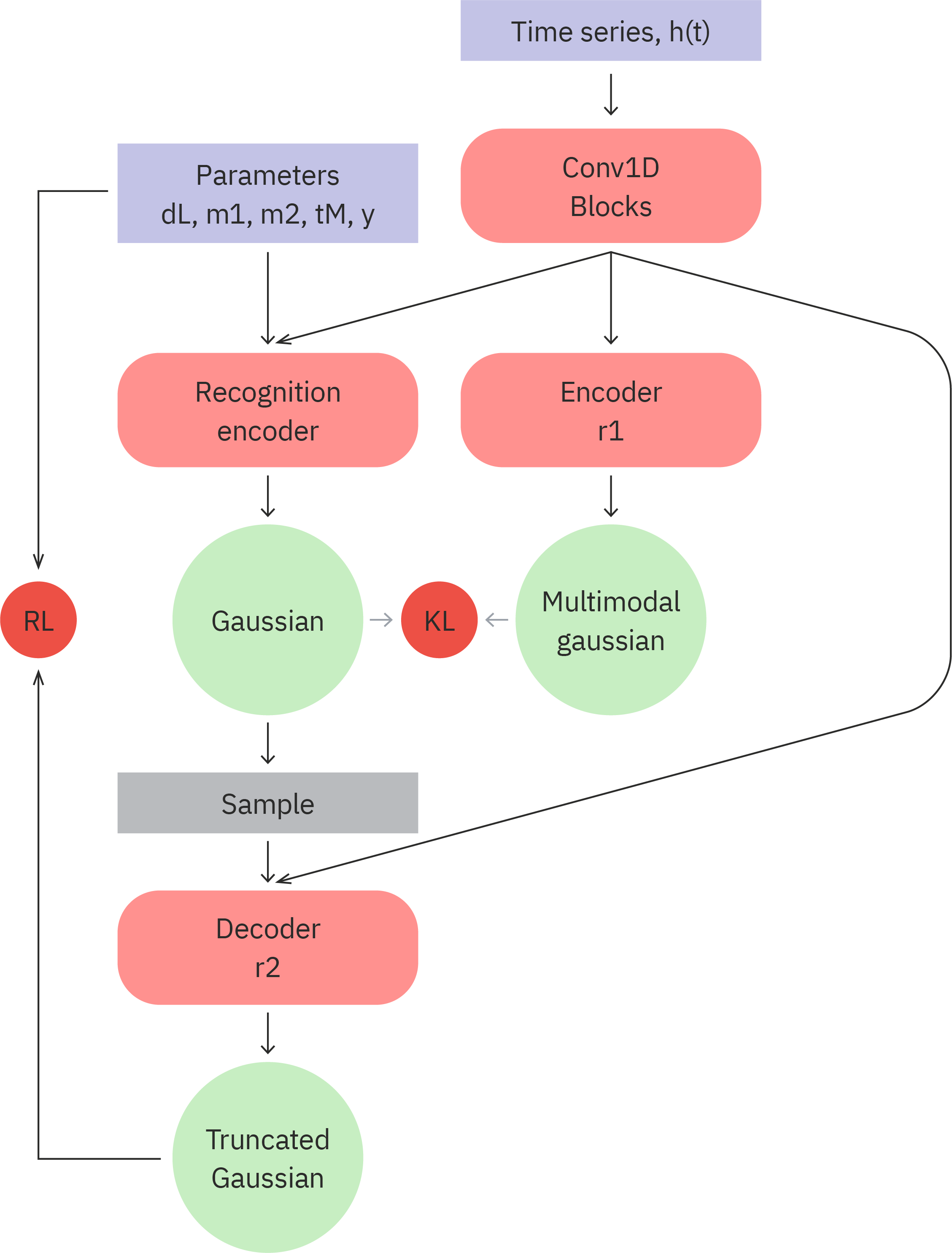}
    \caption{Configuration of the CVAE network for the training phase. More information about the training flow can be found in Appendix \ref{appendix}.}
    \label{fig:train}
\end{figure}

\begin{figure}[t]
    \centering
    \includegraphics[width=0.5\linewidth]{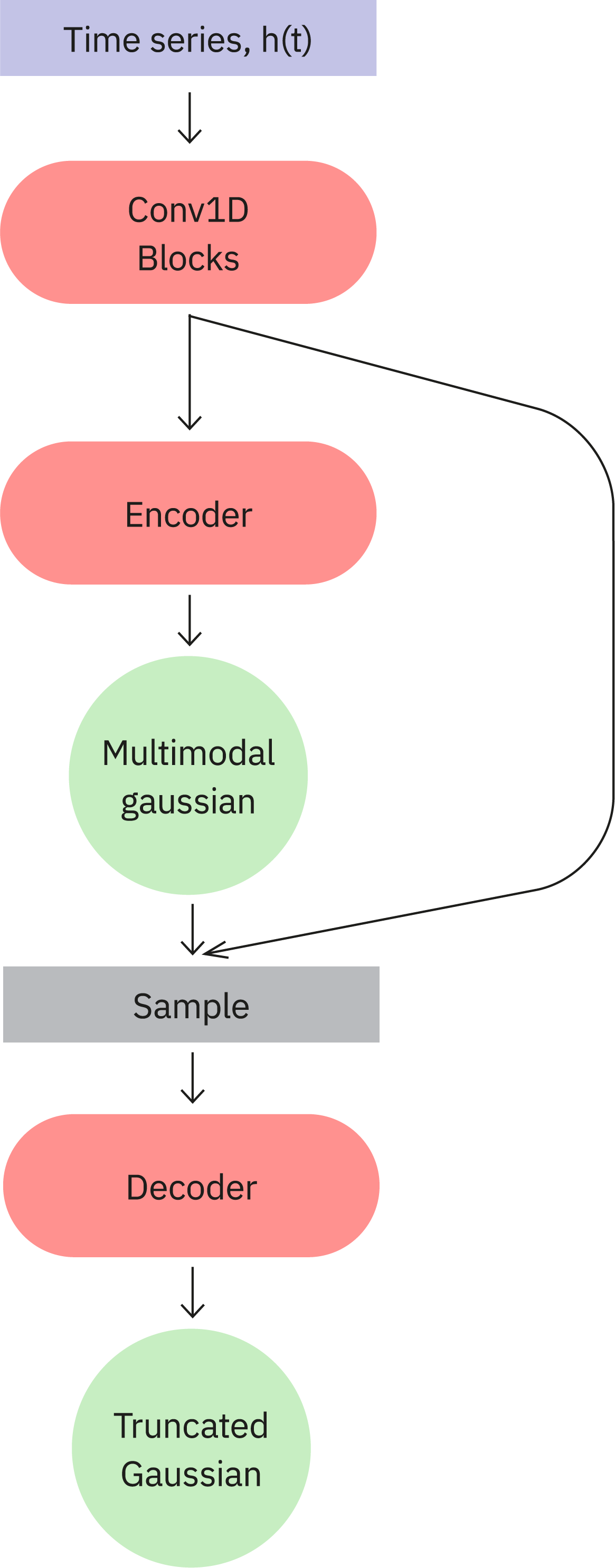}
    \caption{Configuration of the CVAE network for the testing phase. More information about the testing flow can be found in Appendix \ref{appendix}.}
    \label{fig:test}
\end{figure}

In this work, we construct a CVAE following the lines drawn 
in~\cite{Gabbard_2021}, specializing it for the estimation of the two parameters determining the lensing effect of a PML, $\Lambda_L=\{t_M,y\}$.  This CVAE consists on a pair encoder-decoder, $r_{\theta_1}$ and $r_{\theta_2}$, together with a recognition encoder, $q_\phi(z|\Lambda,h)$, where $\theta_1$, $\theta_2$ and $\phi$ represent the set of trainable parameters of each network and $\Lambda=\{d_L,m_1,m_2,t_M,y\}$ are the parameters fed to the recognition encoder. The other GW parameters were not fed to the model in an effort to reduce the number of trainable parameters of the CVAE. The small changes done to the network with respect to the one in \cite{Gabbard_2021} are detailed in Appendix \ref{appendix}, as well as the details concerning the training process. A scheme of the network used for the training phase can be seen in Fig.~\ref{fig:train}. Correspondingly, a scheme of the test phase is shown in Fig.~\ref{fig:test}. Notice that the conditioning information given to the CVAE is precisely the time series of the GW, $h(t)$.

\section{Data}
\label{sec_data}


In this study we rely on utilities from the \textsc{Bilby} software package \cite{bilby_paper} in order to generate both lensed and unlensed BBH merger waveforms. The process of generation is as follows: first, we generate noise data for the Hanford (H1) and Livingston (L1) LIGO detectors at a sampling rate of 2048 Hz, following the methodology described in \cite{data_challenge}. Specifically, we use \textit{Dataset 3}, which consists of purely Gaussian noise, sampled according to power spectral densities calculated from real data from the O3a observation run. Then, we sample the physical parameters of a given waveform from the default \textsc{Bilby} BBH prior values save for the component masses, which are uniformly sampled in the $[35, 80]~M_\odot$ interval and forced to satisfy $m_1>m_2$, following \cite{Gabbard_2021}. We additionally sample the lensing parameters $y\in\left[0.01, 3.0\right]$ and $t_M\in[0.001,0.2]$ seconds, as well as a target signal-to-noise ratio (SNR) in the interval $\textrm{SNR}\in[8,40]$. 
The specified range of $t_M$ corresponds to lens masses within $50 < M_{Lz}/M_\odot < 10^4$.
The priors for the rest of parameters are the default priors used by \textsc{Bilby}. The physical parameters are then passed to the waveform generator, which uses the \texttt{IMRPhenomXPHM} approximant \cite{PhenomXPHM}. This waveform generator computes the transmision factor following the approach explained in Section \ref{sec_pml}. Lensing is applied to the resulting frequency-domain waveform according to 
\beq
    \tilde{h}_L(f; \Lambda) = F(f; \Lambda_L) \cdot \tilde{h}(f; \Lambda_U)\,,
\label{lensed_strain_param}
\eeq 
which is a parametrized version of Eq.~\eqref{lensed_strain}. In this expression $\Lambda_U$ denotes the set of source parameters defining the unlensed waveform, while $\Lambda_L= \{t_M, y\}$ represents the parameters governing lensing effects. Moreover, $\Lambda$ denotes the union of these two sets. This waveform is then injected at the center of a randomly chosen time in an 8-second segment of noise data and the luminosity distance is iteratively adjusted so the SNR of the injected signal matches the target. After this, the data is whitened by inverse spectrum truncation. Since whitening causes artifacts near the edges of the data, we keep only the central 4 seconds for the analysis. The resulting time-series and its physical parameters are then saved to an HDF5 file.



\begin{figure}[t]
    \centering
    \includegraphics[width=1.0\linewidth]{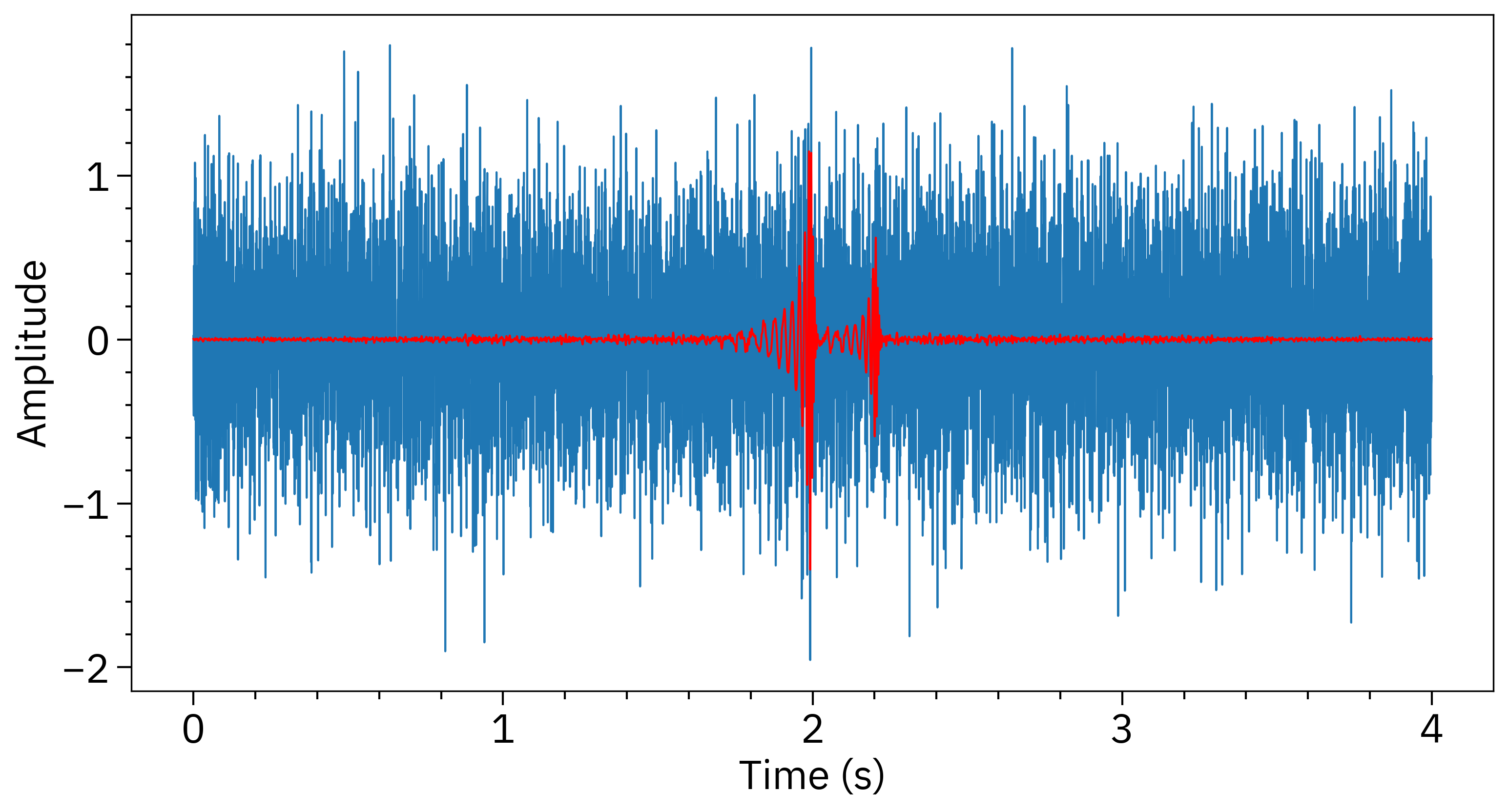}
    \caption{Example of a whitened, lensed BBH waveform from the datasets. The signal is depicted in red, while its noise-masked version appears in blue. For this signal, with parameters: $t_M=0.121\,$s, $y=0.842$, the lensing effect is strong enough to produce two distinct images with an SNR of $25.3$.}
    \label{fig:waveform}
\end{figure}

The sizes of the training, validation and testing datasets are, respectively, $10^6$, $2\times10^4$ and 1000. These datasets do not intersect to ensure that we can monitorize the generalization ability of the model. An example of a whitened, lensed waveform from one of the datasets is shown in Fig.~\ref{fig:waveform}.


\section{Results}
\label{sec_results}

We train our CVAE using the architecture detailed in Appendix~\ref{appendix}, achieving a smooth evolution of the loss function, as shown in Fig.~\ref{fig:loss} (specific definitions are provided in the appendix).  During the initial 100 epochs, while the annealing process is active, the loss shows greater fluctuations. After this period, when annealing ends and the learning rate is set to $10^{-5}$, the loss evolution becomes noticeably smoother.  The loss converges to approximately -2.5, with no signs of overfitting, as both training and validation losses remain close throughout. The training process took between 7 and 8 hours.

\begin{figure}[b]
    \centering
    \includegraphics[width=1.0\linewidth]{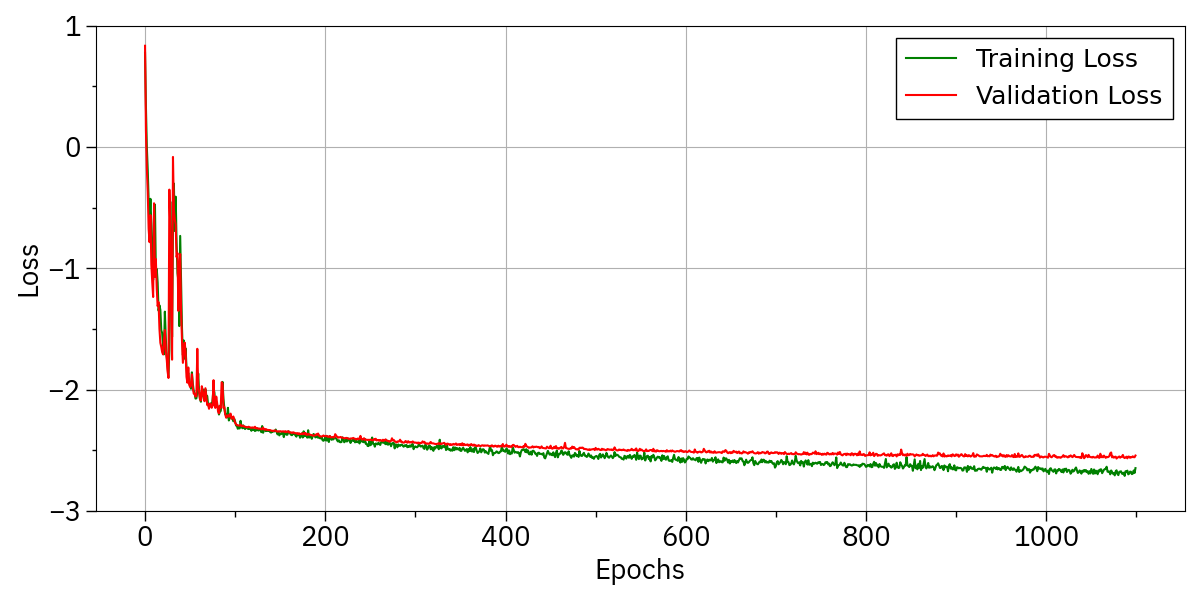}
    \caption{Train and validation loss evolution during training.}
    \label{fig:loss}
\end{figure}

\begin{figure}[t]
    \centering
    \includegraphics[width=1\linewidth]{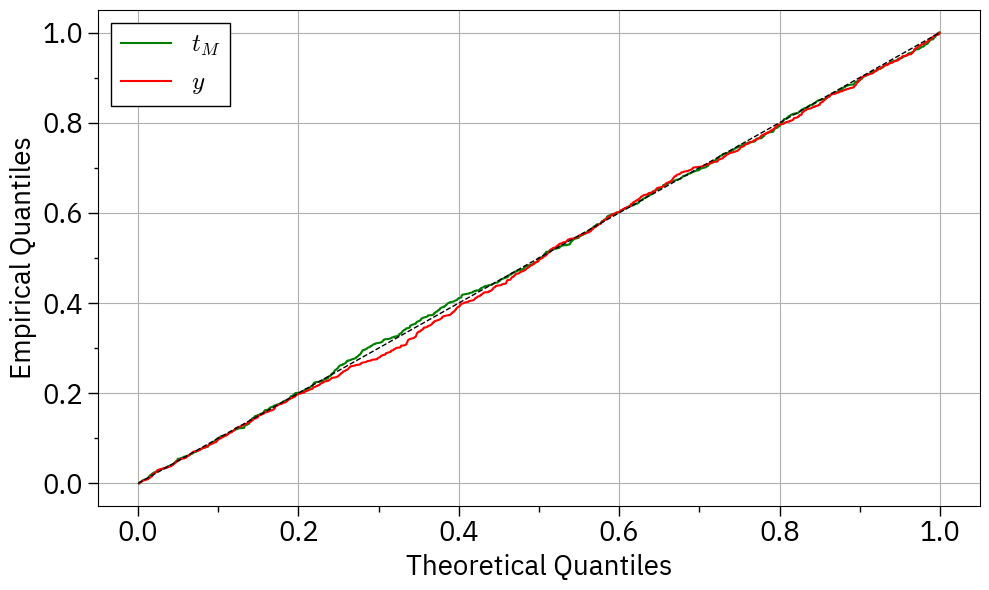}
    \caption{Probability-probability (p-p) plot comparing the posterior probabilities predicted by the CVAE with the true values for the two model parameters. Close alignment with the dashed diagonal line indicates good calibration, meaning that the inferred values fall within the predicted posterior probability intervals as expected.}
    \label{fig:pp}
\end{figure}

At the end of training, we save the model from the epoch with the best performance---specifically, the epoch with the lowest validation loss, which occurred at epoch 1078---and use it for testing.
Using the trained CVAE, we generate 8000 samples for each of the 1000 test waveforms, produced following the guidelines defined in Section \ref{sec_data}. These samples define the posterior distributions for the two parameters estimated by the CVAE, $y$ and $t_M$. A probability-probability (p-p) plot comparing the distribution of parameters of the test set with the distribution of the generated posterior samples is shown in Figure~\ref{fig:pp}. A p-p plot compares the cumulative distribution of predicted probabilities to the cumulative distribution of observed data. In this plot, the $x$-axis represents the quantiles of the cumulative probability predicted by the model (e.g.~the quantile range 0\% to 100\%), whereas the $y$-axis represents the observed frequency of true values falling within these predicted quantile ranges. When the plot lines are close to the diagonal line, it indicates that the model’s predicted probability intervals match the observed data well. In other words, for an $X\%$ quantile on the $x$-axis, approximately $X\%$ of the true values fall within that quantile range in the observed data. A p-p plot close to the diagonal line suggests that the model is well-calibrated: the predicted uncertainty aligns well with the actual uncertainty in the data. As Fig.~\ref{fig:pp} shows, our trained CVAE yields satisfactory results for both model parameters.

\begin{figure}[t]
    \centering
    \includegraphics[width=1\linewidth]{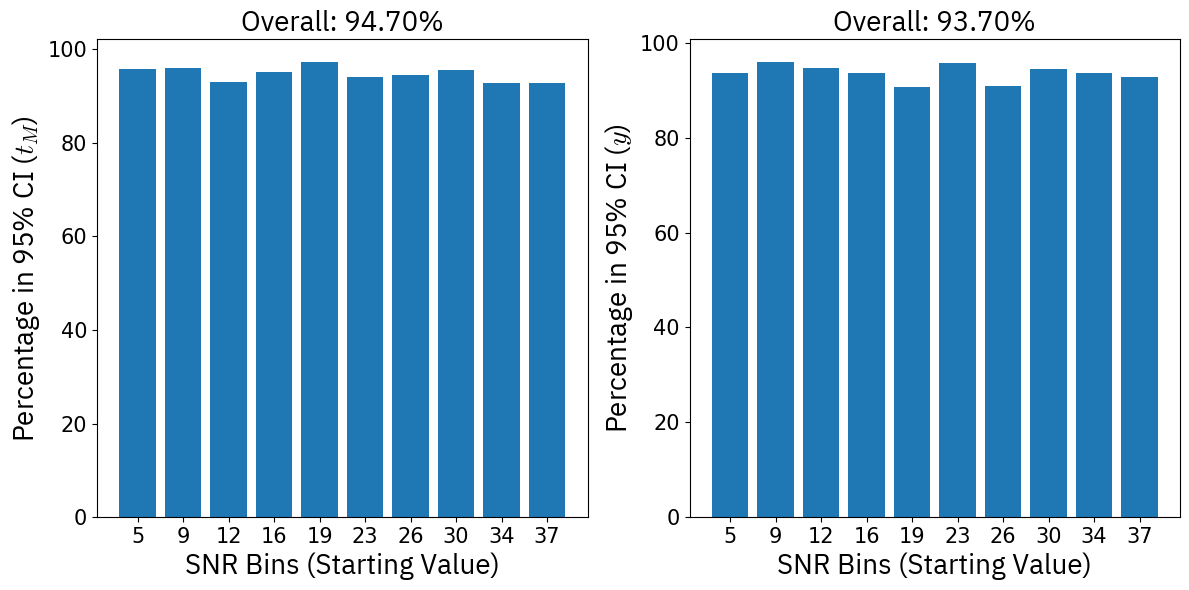}
    \includegraphics[width=1\linewidth]{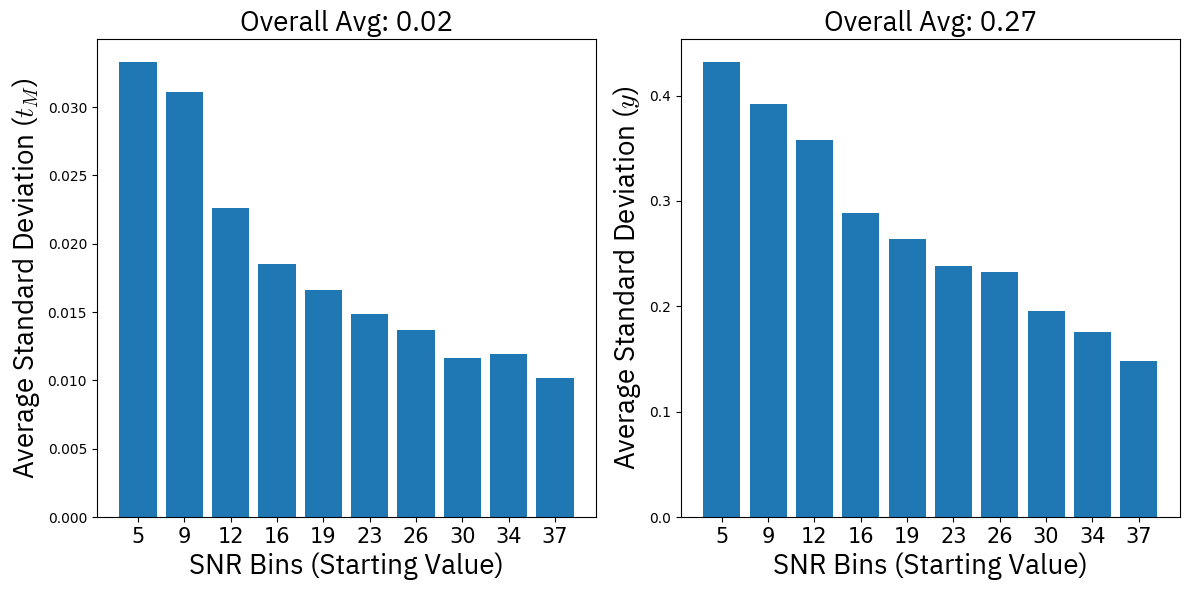}
    \caption{{\it Top panel:} Bar plots showing the percentage of true values of $t_M$ (left) and $y$ (right) inside the 95\% confidence intervals, arranged by SNR. {\it Bottom panel:} same as top panel but showing average standard deviation of the posteriors for $t_M$ (left) and $y$ (right).}
    \label{fig:bar_ci}
\end{figure}

The top panel of Fig.~\ref{fig:bar_ci} shows the percentage of test waveforms that fall within the 95\% confidence intervals of the posterior distributions, arranged by SNR. Notably, these percentages remain consistent across different SNR levels. Approximately 95\% of the true values lie within the 95\% confidence intervals for both $t_M$ and $y$, aligning with the findings in the p-p plot. We also show the average standard deviation of the posterior distributions in the bottom panel of Fig.~\ref{fig:bar_ci}, which is higher for waveforms with lower SNR. Nevertheless, even in the cases of low SNR, the average standard deviations of the posteriors define confidence intervals significantly narrower than the original prior ranges of $y$ and $t_M$.

\begin{figure}[t]
    \centering
    \begin{subfigure}
        \centering
        \includegraphics[width=\linewidth]{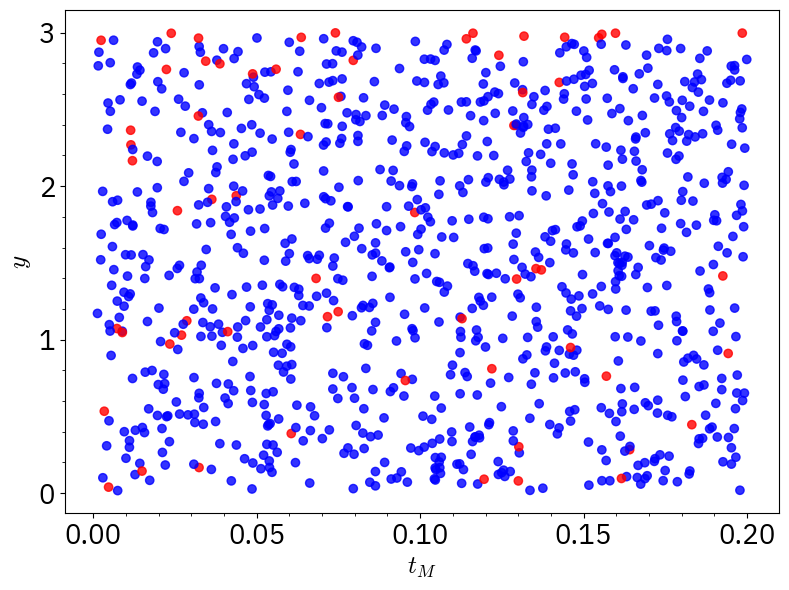}
    \end{subfigure}
    \hfill
    \begin{subfigure}
        \centering
        \includegraphics[width=\linewidth]{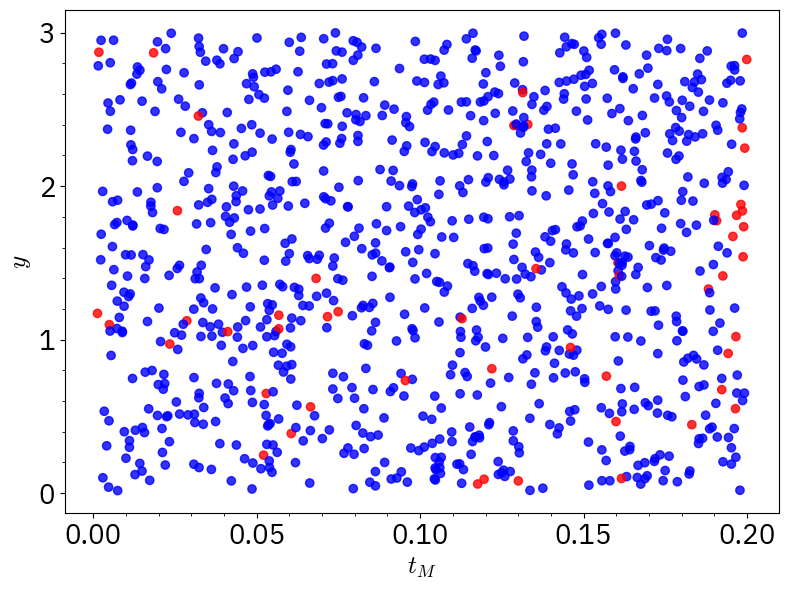}
    \end{subfigure}
    \caption{Scatter plots showing the data points (test set waveforms) based on whether their true values of $y$ (top) or $t_M$ (bottom)  fall within (blue) or outside (red) the 95\% confidence intervals defined by the posteriors.}
    \label{fig:scatter}
\end{figure}

The scatter plots in Figure \ref{fig:scatter} show the test set waveforms, labeled by their values of $y$ and $t_M$. The color code in the dots indicates  whether the waveforms are correctly (blue) or incorrectly (red) processed by the CVAE, i.e., if the true values lie within the 95\% credible intervals defined by the posterior generated by the CVAE. 
Notably, the CVAE successfully estimates $y$ and $t_M$ even when $t_M$ is close to 0 (i.e.~when the lensing effect is nearly absent), demonstrating its ability to handle unlensed waveforms. Additionally, the model accurately estimates the two lensing parameters for values of $y>1$, where the effect is weaker. These cases are challenging to identify, and previous work, such as \cite{kim20}, did not even account for waveforms with those $y$ values. 

A noticeable accumulation of errors at higher values of $y$ and $t_M$ can be observed in the top and bottom panels of Fig.~\ref{fig:scatter}, for $y$ and $t_M$, respectively. The underlying cause of this trend is uncertain and might be a stochastic effect. It is important to note that these plots indicate whether values fall within or outside the 95\% confidence interval, so we expect about 5\% of the waveforms to be incorrectly processed. A deeper investigation focused around the specific range of those high values of $y$ and $t_M$ may be worth pursuing to clarify the source of the observed trend and will be reported elsewhere.

Finally, we analyze the efficiency of the CVAE to perform the inference of the lens parameters, comparing our timing estimates to those of \textsc{Bilby}. In order to do so we perform parameter estimation on 17 different BBH waveforms and measure the inference (wall-clock) time\footnote{The computations were carried out at the ICCUB Nyx's cluster, equipped with dual AMD EPYC 7763 processors, each with 64 cores.}. These estimations were performed using the \textsc{Bilby} package, employing the same waveform generator used to simulate our data.
The results are presented in Table~\ref{tab:time}.
The values reported in this table are obtained from those of Table~\ref{tab:all} in Appendix~\ref{appendix} which shows all timing estimates for the entire set of waveforms. We observe that the CVAE is significantly more efficient than \textsc{Bilby}, achieving speeds  about 4 orders of magnitude faster (occasionally even 5 orders of magnitude; see Table \ref{tab:all}). To accelerate \textsc{Bilby}’s convergence, we use the true values as priors for all parameters except $y$ and $t_M$. For $y$ and $t_M$, we set uniform priors ranging from 0 to 3 and from 0 to 0.2 seconds, respectively. When the CVAE is used to provide priors for \textsc{Bilby}, the priors for these two lensing parameters are uniform distributions defined by the 95\% confidence intervals of the CVAE’s posterior estimates, i.e.~$\mu\pm2\sigma$, where $\mu$ and $\sigma$ represent the posterior mean and standard deviation respectively. In this setup, the average runtime is reduced by 47.9\% compared to the previous configuration.
\begin{table}[t]
    \centering
    \begin{tabular}{c|c|c|c|c}
        \hline
         & 
         {CVAE} & {\textsc{Bilby}} & {\textsc{Bilby}+{\text{CVAE}}} & Saving 
         \\
         & (s) & (s) & (s) & (\%) 
         \\
        \hline
        \hline
        Average & 4 & 54045 & 28159 & 47.9\\
        \hline
        Maximum & 4 & 145063 & 81271 & 79.7\\
        \hline
        Minimum & 4 & 27197 & 15932 & 9.7\\
        \hline
        
    \end{tabular}
    \caption{Comparison of the efficiency between our CVAE and \textsc{Bilby} for parameter estimation. The first two columns report the wall-clock time (in seconds) needed to infer the parameters of a single waveform, showing the longest, shortest, and average durations across 17 runs, for both the CVAE and \textsc{Bilby} in standalone mode. The third column shows the runtime of \textsc{Bilby} when using priors provided by the CVAE. The fourth column presents the percentage of time saved in the three cases.}
    \label{tab:time}
\end{table}

\begin{figure}[t]
    \centering
    \includegraphics[width=1\linewidth]{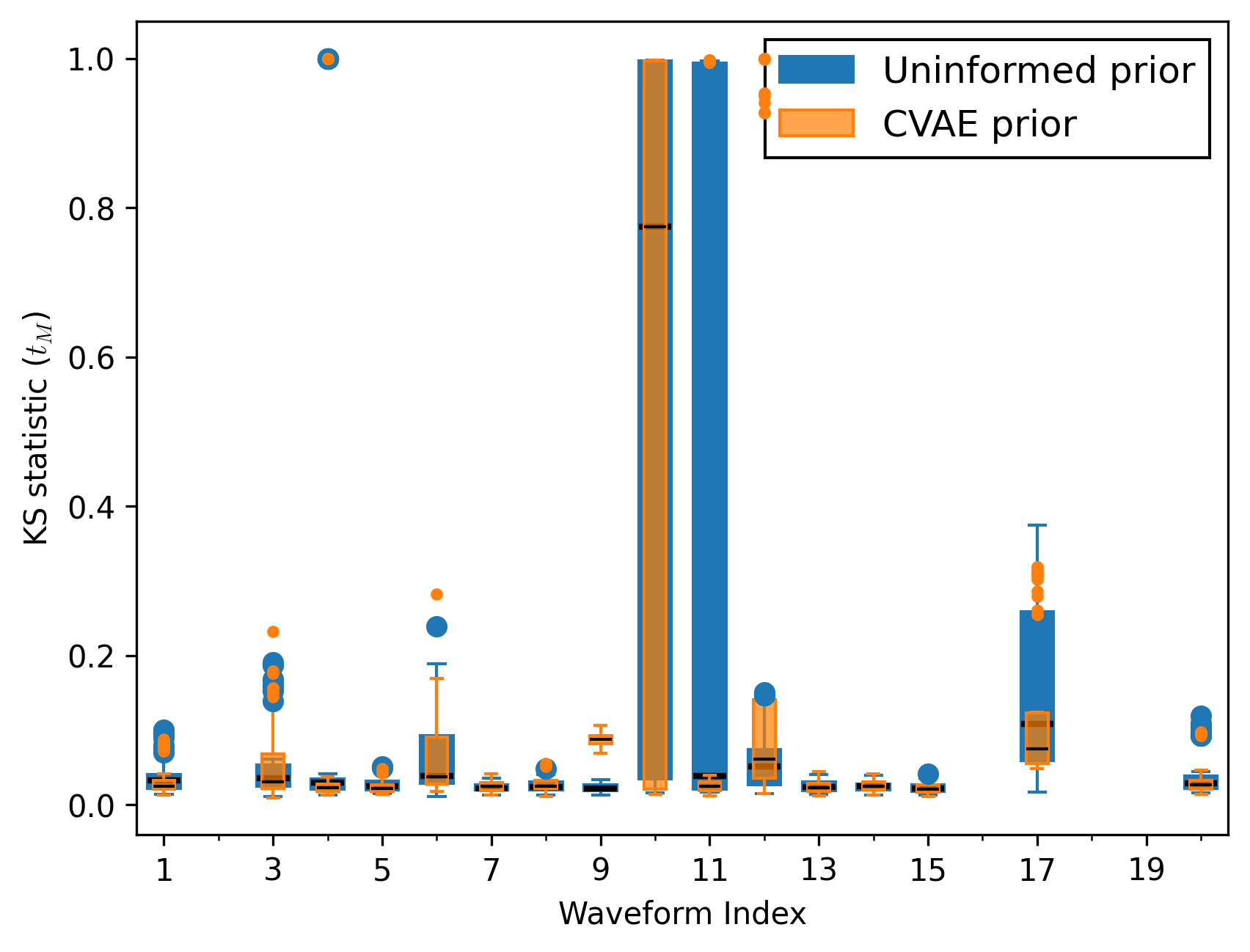}
    \includegraphics[width=1\linewidth]{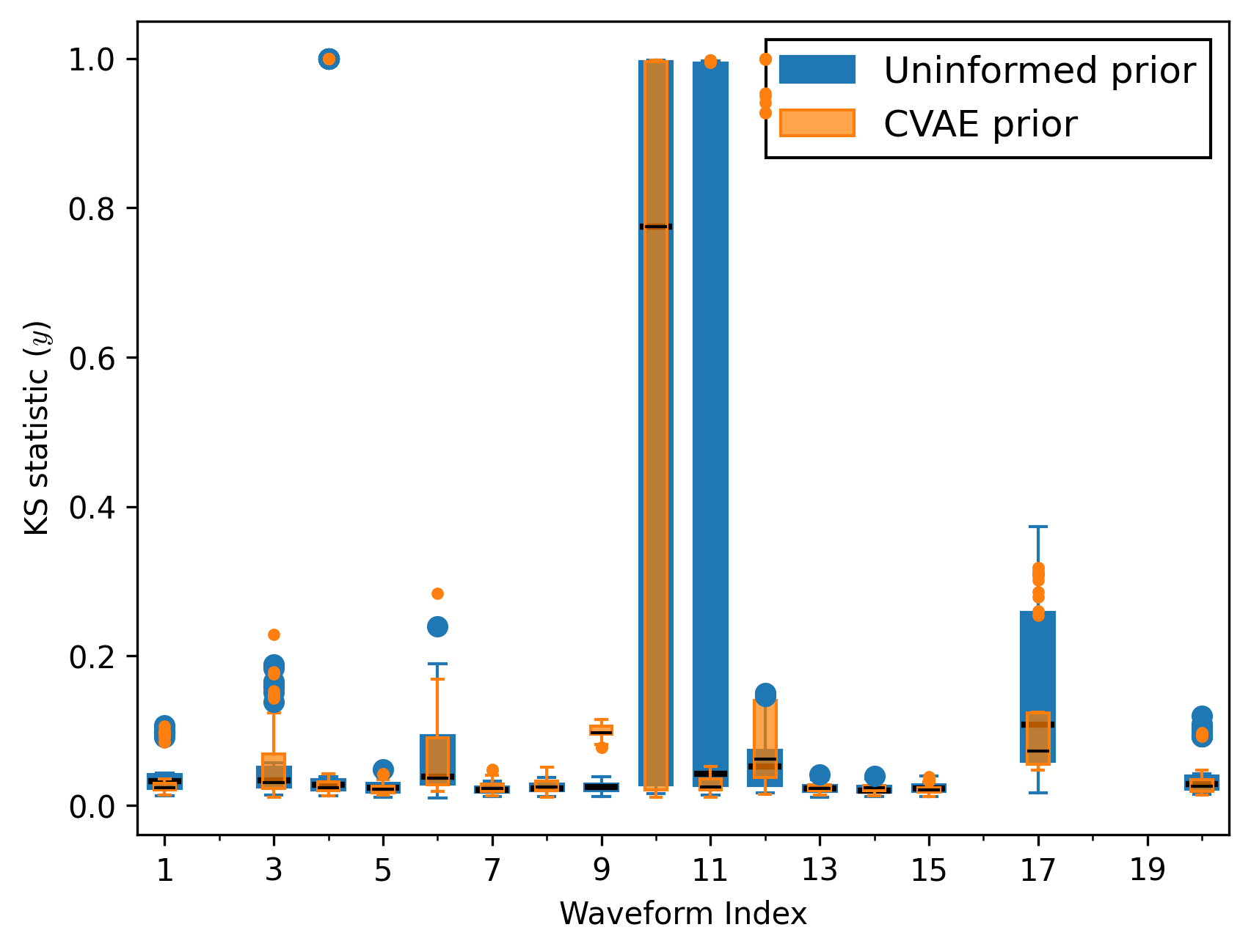}
    \caption{Box plot comparing KS statistics for uninformed\slash uninformed runs (blue) and CVAE-informed\slash uninformed runs (orange) for $t_M$ (top) and $y$ (bottom). The black horizontal lines inside the boxes represent the median value,  the boxes represent the interquartile range (IQR) of the data. The whiskers extend up to 1.5 times the IQR, with values beyond this range considered outliers, shown as individual circles}.
    \label{fig:ks_stat}
\end{figure}

In addition to significantly speeding up the inference, the use of the CVAE combined with \textsc{Bilby} does not degrade the accuracy of the parameter estimation obtained by \textsc{Bilby} when used in standalone mode. 
To demonstrate this, we performed 10 parameter estimation runs for both prior configurations on 20 randomly selected waveforms (indexed 1 to 20). Four of these waveforms (4, 16, 18, and 19) failed to converge in the uninformed runs and were excluded from the analysis; however, we note that the corresponding CVAE-informed runs did converge. For each remaining waveform, we computed the KS statistic between posteriors from different uninformed runs (excluding diagonal elements) and compared it to the KS statistic between uninformed and CVAE-informed runs. If the CVAE-informed prior leads to posterior distributions equivalent to those obtained with an uninformed prior, the KS statistic for CVAE-informed vs. uninformed runs should be comparable to that of different uninformed runs for the same waveform. The results, summarized in the box plots in Fig.~\ref{fig:ks_stat}, largely support this hypothesis: the median KS values for both comparisons lie within one quartile of each other, and outlier values, when they exist, appear in both cases. The only notable exception is waveform 9, where the KS statistic is significantly higher due to the CVAE-informed prior truncating a low-probability tail in the posterior. However, this does not affect the bulk of the distribution. Waveforms 10 and 11 show a large spread in KS statistic values due to noise-induced degeneracy in the matched-filter likelihood. This causes multiple convergence points in the ($tM$,$y$) plane for PE runs on waveforms 10 and 11 which leads to large KS statistics when comparing runs. The use of the CVAE prior restricts the sampling space and can mitigate this effect, as seen in waveform 11.


\section{Conclusions and outlook}
\label{conclusions}

We have studied the suitability of Conditional Variational Autoencoders (CVAE) to conduct  parameter inference of microlensed GWs from binary black hole mergers. Specifically, we have trained a CVAE to estimate the lensing parameters of a point mass lens model, namely $y$ (the relative source position) and $t_M$ (the characteristic lensing time). Our study has shown that CVAE are quite capable to perform this task, yielding high performance in terms of computational efficiency.

The CVAE model has displayed substantial advantages over traditional Bayesian methods in computation time, up to five orders of magnitude faster than \textsc{Bilby}, at the expense of some loss in accuracy. This efficiency makes CVAE an attractive option for real-time or low-latency applications in GW data analysis, addressing one of the main challenges in current microlensing studies, computational cost. Furthermore, our analysis has revealed that incorporating CVAE-generated priors in the inference can further enhance Bayesian frameworks like \textsc{Bilby}. This suggests a promising hybrid approach that combines the rapidness of machine learning and the reliability of traditional statistical inference.

Future research can focus on refining the architecture of the CVAE and making it more suitable for real-world applications. First and foremost, it will be essential to simulate the data using detector noise. In addition, the inclusion of normalizing flows, as proposed by \cite{Green_2020}, may enhance the model's performance. Normalizing flows have proven to be a very powerful tool for parameter estimation in GW science \cite{Williams_2021, Williams_2023, Green_2020}, and hence they can be an important way of improving the model. Furthermore, incorporating spectrograms as extra input data could enable the CVAE to extract additional features present in the time-frequency representation, potentially leading to more precise parameter estimates. Enlarging the number of parameters estimated by the CVAE is also a future goal. Finally, applying the model to real data could serve as a valuable validation step.

\section*{Acknowledgements}
\label{Acknowledgements}
The authors are grateful to Mick Wright and Sreekanth Harikumar for their valuable support and guidance with the  \textsc{Gravelamps} package. 
We also appreciate Narenraju Nagarajan and Melissa Lopez Portilla for their helpful comments on the manuscript. 
Finally, we thank the anonymous referee for their useful and constructive feedback on the first version of this manuscript.
This work was supported by the Spanish Ministry of Science and Innovation (grants PID2021-125485NB-C21 and PID2021-125485NB-C22 funded by MCIN/AEI/10.13039/501100011033 and ERDF, ‘A way of making Europe’), along with the Agència de Gestió d’Ajuts Universitaris i de Recerca (AGAUR), Generalitat de Catalunya (grant SGR-2021-01069), and the Generalitat Valenciana (grant CIPROM/2022/49).
RBN acknowledges financial support from NTT DATA Spain and expresses gratitude to Prof.~Javier Béjar for his guidance on machine learning matters. OGF is supported by an FCT doctoral scholarship (reference UI/BD/154358/2022). Computations have been performed at the Nyx cluster of the ICCUB. This material is based upon work supported by NSF’s LIGO Laboratory which is a major facility fully funded by the National Science Foundation.


\appendix
\section{Features of the CVAE model}
\label{appendix}

We include in this appendix details on the architecture of our CVAE and on the training and testing processes. Our model follows the general lines drawn in \cite{Gabbard_2021} with some minor modifications that we discuss below. The architecture is specified in Table \ref{tab:vae_layers}. Notice that the output shape in the case of the convolutional 1D layers is given by the formula
\begin{equation}
    L_{\rm out}=\bigg\lfloor\frac{L_{\rm in}-\text{Kernel Size}}{\text{Stride}}+1\bigg\rfloor,
\end{equation}
since we use no padding or dilatation. All the layers, except for the last linear layer of the encoder, decoder and recognition networks, have Leaky ReLU \cite{maas2013rectifier} as activation function. The weights of the convolutional block are initialized using the He method at the beginning of the training process. With respect to the CVAE presented in \cite{Gabbard_2021}, minor changes were carried out, most of them concerning the dimensions of the layers. This was motivated by the fact that our input signal is longer and we are estimating only two parameters. We also added an extra convolutional layer in the shared convolutional block. These decisions were taken in part by trial and error, and the final configuration is the one that achieved the best results.

\begin{table*}[!htbp] 
\centering
\begin{tabular}{c|c|c|c|c}
\hline
\textbf{Component} & \textbf{Layer} & \textbf{Input Shape} & \textbf{Output Shape} & \textbf{Parameters} \\ \hline \hline
\multirow{7}{*}{Shared Conv}   & Convolutional & [4096, 2] & [4033, 96] & Kernel Size = 64, Stride = 1 \\                                    \cline{2-5} 
                               & Convolutional & [4033, 96] & [3970, 96] & Kernel Size = 64, Stride = 1 \\ \cline{2-5} 
                               & Convolutional & [3970, 96] & [652, 96] & Kernel Size = 64, Stride = 6 \\ \cline{2-5} 
                               & Convolutional & [652, 96] & [589, 96] & Kernel Size = 64, Stride = 1 \\ \cline{2-5} 
                               & Convolutional & [589, 96] & [140, 96] & Kernel Size = 32, Stride = 4 \\ \cline{2-5} 
                               & Convolutional & [140, 96] & [109, 96] & Kernel Size = 32, Stride = 1 \\ \cline{2-5} 
                               & Convolutional & [109, 96] & [39, 96]  & Kernel Size = 32, Stride = 2, Flatten output $\rightarrow [3744,1]$ \\ \hline
\multirow{4}{*}{Encoder}       & Linear & [3744, 1] & [4096, 1] & - \\ \cline{2-5} 
                               & Linear & [4096, 1] & [2048, 1] & - \\ \cline{2-5} 
                               & Linear & [2048, 1] & [1024, 1] & - \\ \cline{2-5} 
                               & Linear & [1024, 1] & [128, 1] & Reshape output $\rightarrow [2, 32, 2]$\footnote{A multivariate Gaussian with 32 modes. Notice that the dimension of the latent space is 2.} \\ \hline
\multirow{4}{*}{Recognition}   & Linear & [3749, 1] \footnote{The output of the convolutional block concatenated with the vector of real parameters of the gravitational wave.} & [4096, 1] & - \\ \cline{2-5} 
                               & Linear & [4096, 1] & [2048, 1] & - \\ \cline{2-5} 
                               & Linear & [2048, 1] & [1024, 1] & - \\ \cline{2-5} 
                               & Linear & [1024, 1] & [4, 1] & Reshape output $\rightarrow [2, 2]$\footnote{A multivariate Gaussian in the latent space.} \\ \hline
\multirow{4}{*}{Decoder}       & Linear & [3746, 1] \footnote{The output of the convolutional block concatenated with a sample from the latent space defined by the recognition (training) or the encoder (testing) networks.} & [4096, 1] & - \\ \cline{2-5} 
                               & Linear & [4096, 1] & [2048, 1] & - \\ \cline{2-5} 
                               & Linear & [2048, 1] & [1024, 1] & - \\ \cline{2-5} 
                               & Linear & [1024, 1] & [4, 1] & Reshape output $\rightarrow [2, 2]$\footnote{A truncated, multivariate Gaussian as final output for parameter estimation.} \\ \hline
\end{tabular}
\caption{Layer description of the Variational Autoencoder (VAE). The different parts of the CVAE (shared convolutional block, encoder, recognition network and decoder) are disposed as shown in Figure \ref{fig:train}.}
\label{tab:vae_layers}
\end{table*}

The training process starts when the time series is passed through the shared convolutional block. The result is, on the one hand, passed through the enconder network, obtaining a multivariate Gaussian distribution with 32 modes. We call it $\mu_{r_1}$ following the notation of~\cite{Gabbard_2021}. On the other hand, the result is concatenated with five of the source and lens parameters $\Lambda=\{d_L,m_1,m_2,t_M,y\}$ and passed through the recognition network, obtaining another multivariate Gaussian distribution. This is called $\mu_q$. From the latter we sample a vector $z$, which is concatenated with the output of the shared convolutional block and passed through the decoder network, obtaining a third multivariate Gaussian distribution, $\mu_{r_2}$. With these four elements, $\mu_q$, $\mu_{r_1}$, $\mu_{r_2}$ and $z$, we compute the loss function, which is the sum of two terms:

\begin{enumerate}
    \item \textbf{The reconstruction term $(r_2)$:} Computed as the negative log likelihood of the lens parameters $\Lambda_L=\{t_m,y\}$ being a sample of the distribution described by the parameters $\mu_{r_2}$. This term is minimized when there is a high probability of sampling $\Lambda_L$ from $\mu_{r_2}$:
    \begin{equation}
        {\rm RL}\,[\mu_2;\Lambda_L]=-\log(r_2(\Lambda_L))
    \end{equation}
    
    \item \textbf{The KL-divergence:} Computed between the distributions defined by the parameters $\mu_q$ and $\mu_{r_1}$, $q(z|h,\ \Lambda)$ and $r_1(z|\Lambda)$. We can approximate the KL-divergence as
    \begin{equation}
        {\rm KL}\,[q(z|h,\ \Lambda_L)||r_1(z|h)]\approx \log\bigg(\frac{q(z|h,\ \Lambda)}{r_1(z|h)}\bigg)\,.
    \end{equation}
    Since this is equal to $\log(q(z|h,\ \Lambda))-\log(r_1(z|h))$, we just have to compute the log probabilities of $z$ being a sample of both distributions. This term attains its minimum value when the distributions $q$ and $r_1$ look alike.
\end{enumerate}

All the trainable parameters are then updated during back-propagation, driven by this loss function. We apply the same annealing process for the KL divergence and the truncated Gaussian distribution from the output than in \cite{Gabbard_2021}. Both the contribution of said loss term and the range of the truncated Gaussian, are controlled by a parameter $\beta$ that multiplies the KL diverge in the loss function. This parameter increases logarithmically from 0 to 1 between epochs 10 and 20. The lower and upper bounds of the truncated Gaussian vary, respectively, from -10 to 0 and from 11 to 1. In each epoch, the CVAE visits $2\times10^4$ instances, and we perform 1100 epochs. Therefore, the CVAE visits the whole dataset (made out of $10^6$ waveform signals) 22 times. The batch size is set equal to 1500 and the learning rate is set equal to $10^{-4}$ during the first 100 epochs and equal to $10^{-5}$ during the rest of the training process. We use an Adam optimizer with L2 regularization (a weight decay of $10^{-5}$) and a validation set of $2\times10^4$ waveforms.

Once the CVAE is trained we can test it on a test set. We use a set of 1000 waveforms that have never been seen by the model before to test its generalization capability. During the testing process, the recognition encoder is not used. Instead, the time series is passed through the convolutional block and then through the encoder. A sample is taken from the latent space, concatenated with the output of the convolutional block and passed through the decoder, obtaining the parameters of two truncated Gaussian distributions. By passing the same time series $N$ times through the model, one gets $N$ different pairs of truncated gaussians. We take a sample from each of these distributions, obtaining a set of $N$ samples that define a posterior probability that can be represented in a corner plot. When $N=8000$ this process takes, on average, around 4 seconds per waveform. The GPU used for training and testing is an NVIDIA A40.

As a final test, we compare the efficiency of the CVAE with \textsc{Bilby}~\cite{bilby_paper} for parameter estimation. We also study if using the CVAE to provide priors to \textsc{Bilby} helps reducing the computational time of the inference. The results of this analysis is shown in Table \ref{tab:all}.
\begin{table*}[htbp!]
    \centering
    \begin{center}
    \begin{tabular}{c|c|c|c|c|c|c}
        \hline
        {CVAE} & \textsc{Bilby} & \textsc{Bilby} + CVAE & {Time saved} & {SNR} & $y$ & $t_M$  \\
        (s) & (s) & (s) & \% & &  & (s) 
        \\
        \hline
        \hline
        4 & 62374 & 30370 & 51.31 & 18.14 & 0.88 & 0.046 \\
        4 & 36327 & 29156 & 19.74 & 25.84 & 2.97 & 0.072 \\
        4 & 102661 & 36330 & 64.61 & 24.49 & 0.74 & 0.184 \\
        4 & 47179 & 22119 & 53.12 & 28.77 & 1.33 & 0.033 \\
        4 & 27197 & 21376 & 21.40 & 16.88 & 2.48 & 0.129 \\
        4 & 47195 & 18809 & 60.15 & 28.00 & 0.45 & 0.147 \\
        4 & 36930 & 15932 & 56.86 & 38.42 & 1.40 & 0.131 \\
        4 & 145063 & 81271 & 43.98 & 15.28 & 1.86 & 0.002 \\
        4 & 48646 & 29838 & 38.66 & 18.62 & 1.12 & 0.088 \\
        4 & 82371 & 35772 & 56.57 & 8.98 & 0.65 & 0.097 \\
        4 & 34325 & 25982 & 24.31 & 27.66 & 2.16 & 0.092 \\
        4 & 74097 & 33421 & 54.90 & 22.26 & 0.34 & 0.086 \\
        4 & 45966 & 29340 & 36.17 & 13.26 & 0.44 & 0.150 \\
        4 & 69607 & 32147 & 53.82 & 24.31 & 1.10 & 0.042 \\
        4 & 134742 & 27308 & 79.73 & 32.64 & 0.61 & 0.081 \\
        4 & 29511 & 26648 & 9.70 & 10.74 & 2.84 & 0.119 \\
        4 & 29263 & 18671 & 36.20 & 31.37 & 1.48 & 0.077 \\
        \hline
    \end{tabular}
    \end{center}
    \caption{Comparison of inference times (in seconds) for the CVAE and \textsc{Bilby}. Each row represents a different waveform, with the corresponding SNR and $y$ and $t_M$ values}.
    \label{tab:all}
\end{table*}

\newpage

\bibliography{VAE-microL}

\end{document}